\documentclass[showpacs,preprint,aps]{revtex4}
\usepackage{graphicx}
\usepackage{dcolumn}
\usepackage{bm}

\begin{document}

\title{Self-consistent analytical solution of a problem of charge-carrier injection at a conductor/insulator interface}
\author{F. Neumann}
\author{Y.A. Genenko}
\email{yugenen@tgm.tu-darmstadt.de}
\author{C. Melzer}
\author{S.V. Yampolskii}
\author{H. von Seggern}
\affiliation{Institute of Materials Science, Darmstadt University of Technology,
Petersenstra{\ss}e 23, D-64287 Darmstadt, Germany}
\date{\today}


\begin{abstract}
\noindent We present a closed description of the charge carrier
injection process from a conductor into an insulator. Common
injection models are based on single electron descriptions, being
problematic especially once the amount of charge-carriers injected
is large. Accordingly, we developed a model, which incorporates
space charge effects in the description of the injection process.
The challenge of this task is the problem of self-consistency. The
amount of charge-carriers injected per unit time strongly depends
on the energy barrier emerging at the contact, while at the same
time the electrostatic potential generated by the injected charge-
carriers modifies the height of this injection barrier itself. In
our model, self-consistency is obtained by assuming continuity of
the electric displacement and the electrochemical potential all over
the conductor/insulator system. The conductor and the insulator are
properly taken into account by means of their respective density of
state distributions. The electric field distributions are obtained
in a closed analytical form and the resulting current-voltage
characteristics show that the theory embraces injection-limited as
well as bulk-limited charge-carrier transport. Analytical approximations of
these limits are given, revealing physical mechanisms responsible
for the particular current-voltage behavior. In addition, the model
exhibits the crossover between the two limiting cases and determines
the validity of respective approximations. The consequences resulting from our
exactly solvable model are discussed on the basis of a simplified indium
tin oxide/organic semiconductor system.
\end{abstract}

\maketitle

\section{Introduction}

Once a conductor forms contact with an insulator, an energy barrier
is formed between the two materials, which impedes the
charge-carrier injection into the insulator. Although this injection
barrier in general sways charge transport through the
conductor/insulator system, only the two limiting cases of very low
or very high injection barriers are often considered.

For low injection barriers, one expects the contact to be ohmic, meaning that the contact is able to supply more charges per unit time
than the bulk of the insulator can support. In this case, a space-charge region is formed and the electric field at the interface
vanishes \cite{Lambert}. Because excess charge-carriers dominate charge transport in insulators, one observes a space-charge-limited
current (SCLC) density of the form $j\sim V^2/L^3$ (in the absence of charge-carrier traps), where $L$ is the sample thickness
and $V$ is the applied voltage. The current-voltage characteristic (IV-characteristic) is determined by the bulk properties of the
material with no influence of the contact properties \cite{Blom1996APL,Blom1996PRB,Neumann04}.

For high injection barriers, one anticipates the injection rate across the conductor/insulator interface to dominate the IV-characteristic
of the system. The models to describe injection are the
Fowler-Nordheim (FN) tunneling model \cite{Sze} or the Richardson-Schottky (RS) model \cite{Sze,Ashcroft} for thermionic injection.
The FN model describes the charge-carrier injection by tunneling through a triangular barrier into an unbound continuum of states.
The current takes then the temperature independent form:
\begin{equation}
j(V)=B\frac{V^2}{L^2}\exp \left [ -\frac{4 L \sqrt{2m_{eff}}\Delta^{3/2}}{3\hbar e V} \right ],
\label{fowler}
\end{equation}
where $e$ is the elementary charge, $m_{eff}$ being the effective mass of a charge-carrier, $\hbar$ is Planck's constant, $B$ is a
constant and $\Delta$ is the height of the barrier. The RS model on the other hand describes charge injection as a thermally activated hopping over the potential barrier, where barrier
lowering due to the superposition of the external electrostatic potential and the image-charge potential is considered. The RS model
predicts the IV-characteristics to follow:
\begin{equation}
j(V)=CT^2\exp{\left [ -\frac{\Delta}{kT}\right ]}\exp{ \left [ \frac{1}{kT}\sqrt{\frac{eV}{4\pi\epsilon\epsilon_0 L}} \right ]}.
\label{richardson}
\end{equation}
Here, $T$ is the temperature, $k$ is the Boltzmann constant, $C$ represents the effective Richardson constant, $\epsilon$ is the
relative permittivity of the insulator and $\epsilon_0$ is the permittivity of vacuum.

An alternative description of the injection process is given by the
drift-diffusion theory involving electron-electron interaction in a
mean-field approximation \cite{Walker2002,Martin2005}. In this
homogeneous continuum model widely used for the description of
conventional crystalline semiconductors, the steady state current
density exemplarily given for holes in one dimension is:
\begin{equation}
j=e\mu_s p_s(x){\cal F}_s(x)- e Dp_s'(x)=\mbox{const},
\label{drift-diffusion}
\end{equation}
where $\mu_s$ is the hole mobility, $p_s$ is the hole density, ${\cal F}_s$ is the electric field in the insulator, and $D$ is
the diffusivity of holes. ${\cal F}_s$ and $p_s$ are coupled by the Poisson equation. The drift-diffusion equation in combination
with the Poisson equation involves space-charge effects, but meets the problem of self-consistency in the boundary conditions. The
electrostatic potential generated by the injected charge-carriers modifies the injection barrier, but on the other hand the amount
of charge-carriers injected per unit time depends strongly on the barrier height. Therefore, a proper description of the injection
process by means of the drift-diffusion theory has to involve both, the insulator and the conductor side of the system.
This can be seen in analogy to a strongly asymmetric {\itshape pn}-junction.

Different attempts to numerically describe the injection process in
insulators with account of the space charge were made recently
\cite{Tutis2001,Neumann2006,Christen2007}. In transitive
FE-simulations in one- and two-dimensional geometries by Christen et
al. \cite{Christen2007} both injection-limited and space
charge-limited regimes were presented. However, the injection
description was still not self-consistent imposing some
phenomenological values for the field and the carrier density at the
contact. A comprehensive one-dimensional numerical model accounting
for space-charge effect was developed by Tuti\v{s} et al.
\cite{Tutis2001} which comprised the hopping transport and the
tunnel injection from electrodes. Unfortunately, this sophisticated
numerical tool does not allow analytical fitting of current-voltage
characteristics which gives insight in major mechanisms controlling
injection. In the work of Neumann et al. \cite{Neumann2006}, a
self-consistent numerical treatment of the injection and transport
processes in a conductor/insulator/conductor device was presented in
which continuity of the electrochemical potential and the electric
displacement was assumed everywhere in the system, in particular at
the contacts. The conductor and the insulator were characterized by
their specific density of state (DOS) distributions. In present work 
we obtain  an exact analytical solution for the injection  across the conductor/insulator
interface in a self-consistent manner. To focus on the consequences
of the self-consistent treatment, we discuss our results on the
basis of a simplified indium tin oxide/organic semiconductor system
used as hole-injecting contact in organic optoelectronic devices,
shifting the comparison with experimental data to a future time.

\section{The model}
Let us consider a single conductor/insulator interface located at a position $x=0$. The conductor is supposed to extend over
the semi-space with $x<0$, whereas the insulator covers the semi-space with $x>0$. All energies are measured downwards with
respect to the top of the valence band in the conductor, in order to account for hole transport in the system. In the
following sections, the theoretical models describing the insulator and the conductor are introduced.

\subsection{Electrode}
The conductor electrode is characterized by its DOS as a function of the energy $E$. For our purpose it is sufficient to assume the
free electron approximation, in which the DOS function reads:
\begin{equation}
g_c(E)=\frac{1}{3\pi^2}\left ( \frac{2m_{eff}}{\hbar^2} \right ) ^{\frac{3}{2}}\sqrt{E},
\label{dosmetal}
\end{equation}
where  $m_{eff}$ is the effective mass in the conductor. In the Thomas-Fermi approximation \cite{Ashcroft}, one can deduce the
electrochemical potential $\kappa_c$ of the conductor as a function of the spatial coordinate $x$,
\begin{equation}
\kappa_c(x)=\frac{\hbar^2}{2m_{eff}}(3\pi^2p_c(x))^{2/3}+e\phi(x),
\label{kappametal}
\end{equation}
where $p_c(x)$ is the hole density in the electrode and $\phi(x)$ is the electrostatic potential. In general, the electrochemical
potential $\kappa(x)$ relates the steady state current density $j$ with the charge-carrier density. For a one-dimensional
geometry, the current remains constant across the whole space and $j$ is given by the conductivity $\sigma$ and the derivative of
$\kappa(x)$ \cite{Landau},
\begin{equation}
j=-\frac{\sigma}{e}\frac{d\kappa(x)}{dx}
\label{currentgeneral}
\end{equation}
(notice the direction of the energy axis). The conductivity of a conductor $\sigma_c=e \mu_c p_\infty$ can be expressed in terms
of the hole mobility $\mu_c$ and the hole density $p_\infty$ in the valence band at an infinite distance from the
conductor/insulator interface.

Since charge-carriers are transferred from the electrode to the insulator, a space-charge region emerges near the interface
which modifies the electric field ${\cal F}_c(x)$ in the conductor according to Gauss law,
\begin{equation}
{\cal F}_c'(x)=\frac{e}{\epsilon_c\epsilon_0}\delta p(x),
\label{gaussmetal}
\end{equation}
where $\epsilon_c$ is the relative permittivity of the electrode.
In Eq.(\ref{gaussmetal}) $\delta p(x)=p_c(x)-p_\infty $ is the excess hole density. However, charge-carrier densities in degenerate
conductors are rather high and consequently, the value for the excess hole density is small in comparison with the background
hole density, $|\delta p(x)|\ll p_\infty$. Hence, the well-known linearized Thomas-Fermi approximation \cite{Ashcroft} can be
applied, leading with Eqs.(\ref{kappametal}-\ref{gaussmetal}) to a differential equation for ${\cal F}_c$:
\begin{equation}
l_{TF}^2{\cal F}_c''(x) - {\cal F}_c(x)= -\frac{j}{\sigma_c}
\label{dglmetal}
\end{equation}
with
\begin{equation}
l_{TF}=\sqrt{\frac{2}{3}\frac{\epsilon_0\epsilon_c \kappa_\infty}{e^2p_\infty}},
\label{ld}
\end{equation}
being the Thomas-Fermi screening length, defining the typical length scale of the system. Here $\kappa_\infty$ is the electrochemical
potential at an infinite distance from the conductor/insulator interface.

Since space-charge zones in conductors are of finite thickness, gradients of ${\cal F}_c(x)$ have to vanish at an infinite distance from the
contact. Therefore, the solution for the electric field reads
\begin{equation}
{\cal F}_c(x)=\left [ {\cal F}_c(0)-\frac{j}{\sigma_{c}} \right ] e^{x/l_{TF}} + \frac{j}{\sigma_{c}},
\label{solutionmetal}
\end{equation}
where the electric field in the conductor at the conductor/insulator interface, ${\cal F}_c(0)$, is the only unknown quantity.

The validity of the used approximations has to be reviewed critically. An exact quantum-mechanical theory of the inhomogeneous
electron gas accounting for the ionic lattice of the material and electron correlations, gives a comparable scale for the electric
field penetration in conductors and demonstrates the usability of the uniform positive background model for simple metals
\cite{Lang1,Lang2}. While in typical metals the Thomas-Fermi screening length is about $1\mathring{A}$ making the application of
the quasi-classical Thomas-Fermi approximation, Eq.(\ref{kappametal}), inappropriate, substantially larger Thomas-Fermi screening
lengths can be found in degenerate semiconductors such as transparent conducting oxides used in optoelectronic devices
\cite{Mergel2002,Mergel2004,Fujiwara2005}.

\subsection{The insulator}
Similarly, the insulator is characterized by a DOS function, $g_s(E)$, describing extended states in which charge transport takes
place. In contrast to the conductor the electrochemical potential, $\kappa_s$, is situated well above the top of the band of
extended hole states. In other words, the insulator is supposed to be non-degenerate. Introducing
a band-edge means that the DOS function $g_s(E)=0$ when $E<0$. Hence, the density of holes in extended
states can be calculated using Boltzmann statistics:
\begin{equation}
p_s(x)=\int\limits_{-\infty}^\infty g_s(E-\Delta-\kappa_\infty)\exp \left ( \frac{\kappa_s(x)-e\phi(x)-E}{kT} \right ) dE,
\label{nsboltzmann}
\end{equation}
where the energy scale has been adjusted to the top of the valence band in the conductor. The injection barrier $\Delta$ is
defined as the energetic difference of the top of the extended state distribution, $g_s(E)$, to the electrochemical potential in
the conductor at an infinite distance from the interface, $\kappa_\infty$. The electrochemical potential of the insulator is
then given by
\begin{equation}
\kappa_{s}(x)=kT\ln \left ( \frac{p_s(x)}{{\cal N}} \right ) +\Delta+\kappa_\infty+e\phi(x),
\label{kappaorganic}
\end{equation}
where the quantity
\begin{equation}
{\cal N}=\int\limits_{-\infty}^\infty g_s(E)\exp(-E/kT) dE
\label{neff}
\end{equation}
can be understood as the effective total density of states available in the insulator at a given temperature. We note that the
temperature dependence of ${\cal N}$ becomes weak in the case of a narrow-band insulator.

Since band-gap energies are much larger than $kT$, thermal excitation of a charge carrier from the valence band to the conduction
band of a typical insulator is virtually impossible. Charge-carriers contributing to the electrical current are therefore excess
charge-carriers injected from the conductor and their total density has to appear in Gauss law,
\begin{equation}
{\cal F}_s'(x)=\frac{e}{\epsilon_s\epsilon_0}p_s(x),
\label{gaussorganic}
\end{equation}
with $\epsilon_s$ the relative dielectric permittivity of the insulator.
Eqs.(\ref{currentgeneral}),(\ref{kappaorganic}) and (\ref{gaussorganic})  with $\sigma_s(x)=e\mu_s p_s(x)$ lead to a
nonlinear differential equation for the electric field ${\cal F}_s(x)$,
\begin{equation}
\frac{kT}{e}{\cal F}_s''(x)-{\cal F}_s(x){\cal F}_s'(x)=-\frac{j}{\mu_s\epsilon_s\epsilon_0},
\label{dglorganic}
\end{equation}
where $\mu_s$ is the hole mobility in the insulator. The same result is obtained by employing the drift-diffusion
model, Eq.(\ref{drift-diffusion}), and the Einstein relation, relating $\mu_s$ and $D$ \cite{Ashcroft}.
Introducing the following dimensionless quantities,
\begin{eqnarray}
X&=&\frac{1}{l_{TF}}x,
\label{xnull}\\
F_s&=&\frac{el_{TF}}{kT}{\cal F}_s,
\label{enull}\\
\iota &=& \frac{e^2l_{TF}^3}{\mu_s\epsilon_s \epsilon_0 (kT)^2}  j,
\label{jnull}
\end{eqnarray}
Eq.(\ref{dglorganic}) converts into a dimensionless form,
\begin{equation}
F_s^{\prime\prime}(X)-F_s^{\prime}(X)F_s(X)+\iota=0.
\label{dglorgdimless}
\end{equation}

The solution of Eq. (\ref{dglorgdimless}) must be separately formulated for the two cases of thermal equilibrium
and an applied steady-state current.
In equilibrium, the dimensionless current density $\iota$ vanishes and Eq.(\ref{dglorgdimless}) can be integrated in elementary
functions. The first integration results in:
\begin{equation}
F_s'( X ) - \frac{1}{2} F_s^2( X ) = \Lambda
\label{dglorglambda}
\end{equation}
\noindent with an arbitrary constant $\Lambda$. At an infinite distance from the contact the field and its derivative (i.e. the charge
carrier density) vanish, thus, $\Lambda=0$ and a solution for the electric field reads,
\begin{equation}
F_s( X )=\frac{F_s(0)}{1-F_s(0) X /2}.
\label{orgequiinf}
\end{equation}
In Eq. (\ref{orgequiinf}) the field at the contact in the insulator, $F_s(0)$, is the only unknown quantity.

Considering a net current density, a general solution of Eq.(\ref{dglorgdimless}) is known in terms of Airy functions
$\mathrm{Ai}$ and $\mathrm{Bi}$ \cite{Polyanin,Abramowitz}:
\begin{equation}
F_s(X)=-2^{2/3}\iota^{1/3}\frac{\mathrm{Ai}^{\prime}\left[(\iota/2)^{1/3}(X+C_1)\right]+
C_2 \mathrm{Bi}^{\prime}\left[(\iota/2)^{1/3}(X+C_1)\right]}
{\mathrm{Ai}\left[(\iota/2)^{1/3}(X+C_1)\right]+
C_2 \mathrm{Bi}\left[(\iota/2)^{1/3}(X+C_1)\right]},
\label{solexact}
\end{equation}
\noindent where primes denote derivatives of Airy functions with respect to their arguments,
$C_1$ and $C_2$ are unknown constants.
When $X\rightarrow\infty$ the gradient of the charge carrier density has to vanish so that,
\begin{equation}
F_s(X) F'_s (X) = \iota.
\label{bcinfinity}
\end{equation}
Eq.(\ref{bcinfinity}) demonstrates that, in presence of a constant current, the magnitude of the electric field $F_s(X)$
rises asymptotically
since the charge carrier density $\sim F'_s(X)$ vanishes. Considering the asymptotic behavior of the solution of
Eq. (\ref{dglorgdimless}) it is convenient to account explicitly for the sign of the current density $\iota$. Since we assume
injection of holes from the left semi-space to the right one, $\iota$ together with $j$ must be positive. This imposes
asymptotes of the field and the hole density
\begin{equation}
F_s(X)\sim \sqrt{X},\,\,p_s\sim 1/\sqrt{X},
\label{asymptotics}
\end{equation}
\noindent respectively, resembling the characteristic behavior for SCLC \cite{Lambert}.
Taking into account asymptotic properties of Airy functions \cite{Abramowitz} it is easy to establish that, to satisfy
the asymptotic conditions for the field, Eqs. (\ref{bcinfinity}) and (\ref{asymptotics}), the constant $C_2$ must equal zero.
Then, the general solution, Eq. (\ref{solexact}), transforms to
\begin{equation}
F_s(X)=-2(\iota/2)^{1/3}\frac{\mathrm{Ai}^{\prime}\left[(\iota/2)^{1/3}(X+C_1)\right]}
{\mathrm{Ai}\left[(\iota/2)^{1/3}(X+C_1)\right]}.
\label{solspecial}
\end{equation}
The constant $C_1$ must be determined from the boundary conditions derived in the next section.

\subsection{Self-consistency and boundary conditions at the contact}
The equations of the electric field distributions in the conductor and in the insulator
 have to be solved self-consistently. Assuming steady-state, self-consistency is achieved by adjusting the integration constants (i.e.
$F_s(0)$, $F_c(0)$ and $C_1$)
with respect to the Maxwell equations and continuity of the electrochemical potential:
\begin{eqnarray}
\label{bcgeneralfield}
\epsilon {\cal F}(x)&=&continuous,\\
\kappa(x)&=&continuous.
\label{bcgeneralkappa}
\end{eqnarray}
Eq.(\ref{bcgeneralfield}) requires that the electric displacement in
the insulator at the interface, $\epsilon_s F_s(0)$, is equal to
$\epsilon_c F_c(0)$. This holds as long as no interface charges or
dipole-layer exist at the contact, being problematic in many systems
considered here \cite{Ishii1999,Veenstra2000,Peisert2002}. Yet, for
simplicity and without loss of generality Eq.(\ref{bcgeneralfield})
is assumed, leading with Eq.(\ref{bcgeneralkappa}) and
Eq.(\ref{solutionmetal}) to a nontrivial boundary condition relating
the electric field and its first derivative in the respective media
at the contact. In the insulator the following boundary condition
holds:
\begin{equation}
\frac{\epsilon_s}{\epsilon_c} F_s(0) - \frac{\Delta}{kT} - \ln(\alpha F'_s(0)) =
\iota \alpha \frac{\mu_s}{\mu_c} \frac{{\cal N}}{p_\infty},
\label{bcssx0}
\end{equation}
where the dimensionless constant $\alpha$ is defined as
\begin{equation}
\alpha=\frac{\epsilon_s\epsilon_0 kT}{\epsilon_c e^2l_{TF}^2{\cal N}}=
\frac{3}{2} \frac{\epsilon_s }{\epsilon_c}\frac{kT}{\kappa_\infty}\frac{p_\infty}{{\cal N}}.
\label{alpha}
\end{equation}
This boundary condition contains parameters of both media, being specified in the bulk of the respective material.

\section{Physical and numerical analysis}
In this section the solution of the injection problem using the presented model is discussed. We distinguish between the
equilibrium condition, where space charge zones are formed as a consequence of diffusive charge carrier transfer, and the case
of a steady state current, where charge carriers are driven through the system by a time invariant external electric
field.

As an example for an insulator, an organic semiconductor can be
considered. Organic semiconductors show many typical characteristics
of insulators like relatively large band gaps up to $3$ eV and
hence, the absence of intrinsic charge carriers. However, it is well
established that disordered organic semiconductors possess a
Gaussian DOS \cite{Baessler1993} compromising the applicability of
Eq.(\ref{nsboltzmann}) - organic semiconductors are degenerate
systems, the tail states acting as charge carrier traps. As a
consequence, one has to distinguish between trap states and
transport states and Fermi statistics has to be considered. Yet, for
weak disorder, the Gaussian width is small and in the limiting case
of a vanishing disorder, charge-carrier trapping in tail states is
negligible and Boltzmann statistics is valid.

In organic light emitting diodes (OLEDs) or field effect transistors (OFETs) organic semiconductors are contacted with
metals like Au, Ca, Al or transparent conducting oxides like indium tin oxide (ITO) to allow for charge carrier injection
in the otherwise charge-carrier free organic semiconductor. While in metals the Thomas-Fermi approximation is disputable
due to the prevailed low screening length of $1\mathring{A}$, the characteristic scale $l_{TF}$ varies in ITO, depending on
doping \cite{Mergel2002,Mergel2004,Fujiwara2005}, from 2.4$\mathring{A}$ to a few $\mbox{nm}$. Therefore, we choose in our
self-consistent consideration an ITO electrode and assume that it can be described in terms of the Thomas-Fermi approximation.
ITO is typically employed as anode in OLEDs \cite{Martin2005}, since it provides a decent conductivity and a sufficient high
workfunction (5eV) to allow for efficient hole injection while being transparent in the visible range of the optical spectrum
to enable light outcoupling.

Due to the importance of charge carrier injection for the device performance of OLEDs and OFETs,
the description of the involved charge carrier injection process has been advanced in recent years
considering surface recombination of charge carriers at the interface \cite{Walker2002,Martin2005,Scott1999,Malliaras1999}
or stochastic hopping in a surface-barrier potential \cite{Arkhipov1998,Arkhipov1999,Arkhipov2003}. Even more suitable for organic
semiconductors microscopic models account for the mobility of electrons within and transfer between molecular strands and for the
interaction of electrons with molecular vibrational modes \cite{Stoneham2002}. However, all these models work within the single
electron picture, so far it concerns the injection process, which means that interaction between injected electrons is not
incorporated and, thus, space-charge effects on the injection are not properly taken into account.

From now on it is assumed that the material specific quantities of the organic semiconductor and ITO adopt the typical values given
in Table \ref{Materialparameters}.
\begin{table}
  \centering
\begin{tabular}{|c|c|c|c|c|c|c|c|c|}
  \cline{1-9}
    \multicolumn{3}{|c|}{organic} &
    \multicolumn{6}{|c|}{ITO} \\
  \hline
  $\cal N$ & $\epsilon_s$ & $\mu_s$ & $p_\infty$ & $m_{eff}$ & $\kappa_\infty$ & $\epsilon_c$ & $\mu_c$ & $l_{TF}$ \\
  \small(cm$^{-3}$) &   & \small($\mbox{cm}^2/\mbox{Vs}$) & \small(cm$^{-3}$) & \small($m_e$) & \small($eV$) &  &
  \small($\mbox{cm}^2/\mbox{Vs}$) & \small($\mathring{A}$)  \\
  \hline
  $10^{21}$ & $3$ & $10^{-4}$ & $10^{20}$ & $0.35$ & $0.225$ & $9.3$ & $30$ & 8.6 \\
  \hline
\end{tabular}
\caption{{\itshape \small Typical material parameters for an organic semiconductor and ITO
\cite{Mergel2002,Mergel2004,Fujiwara2005}. The parameters are deduced assuming $T=300K$. $m_e$ is the electron mass and
$\alpha=5.6\cdot 10^{-3}$ is determined by parameters of both materials.}}
\label{Materialparameters}
\end{table}
Thereby, the injection barrier $\Delta$ is given by the energetic difference between $\kappa_\infty$ and the band
edge in the organic semiconductor and, thus, is determined by both media. Changing the value of the barrier height while leaving
the electrode unchanged can therefore be understood as considering a different organic semiconductor.

Specifying the material parameters for the insulator and the conductor leads to substantial consequences for the boundary condition
given in Eq.(\ref{bcssx0}). $\iota$ is multiplied  by a small factor, $\alpha (\mu_s/\mu_c) ({\cal{N}}/p_\infty) \sim 10^{-7}$ and hence,
the boundary condition does not depend directly on $\iota$ in most practical cases. Neglecting $\iota \alpha (\mu_s/\mu_c) ({\cal{N}}/p_\infty)$,
Eq.(\ref{bcssx0}) can be reformulated to
\begin{equation}
p_s(0)={\cal N} \exp\left[(\epsilon_s/\epsilon_c) F_s(0) - \Delta/kT\right],
\label{nnull}
\end{equation}
$p_s(0)$ being the density of injected holes at the contact. As a consequence, the dependence of $F_s(0)$ on the current is only due to
Eq.(\ref{dglorgdimless}). Apparently, there exists a barrier
variation $\propto F_s(0)$ leading to the definition of an effective injection barrier $\Delta_{\mbox{eff}}$ of
\begin{eqnarray}
\label{deltaeff}
\Delta_{\mbox{eff}}&=&\Delta - k T(\epsilon_s/\epsilon_c) F_s(0)\\
&=&\Delta - e{\cal F}_c(0)l_{TF}.
\label{deltaeffdim}
\end{eqnarray}
The modification of the injection barrier corresponds to the amount of energy a charge carrier gains (or loses) in the electric
field at the electrode side of the interface. This change, as is seen from Eq.(\ref{deltaeffdim}), may be positive or negative,
since the space-charge region at the interface represents a potential barrier itself.

\subsection{Equilibrium}

Using the analytical solution (\ref{orgequiinf}) valid for thermal equilibrium, the boundary condition (\ref{bcssx0}) can be simplified to a transcendental
relation for $F_s(0)$:
\begin{equation}
\frac{\epsilon_s}{\epsilon_c} F_s(0)-\frac{\Delta}{k T}-\ln \left ( \frac{\alpha}{2} F_s^2(0) \right ) =0.
\label{bcequi}
\end{equation}
$F_s(0)$ together with Eq.(\ref{orgequiinf}) yields the solution of the electric field
distribution in thermal equilibrium. While the charge carrier mobilities of the respective material determine the time
needed to reach equilibrium they apparently do not influence the final equilibrium electric field distribution.
Eq.(\ref{orgequiinf}) and Eq.(\ref{bcequi}) are generally valid for non-degenerate systems.

\begin{figure}[h]
\begin{center}
\includegraphics[height=6.4cm,width=7.5cm]{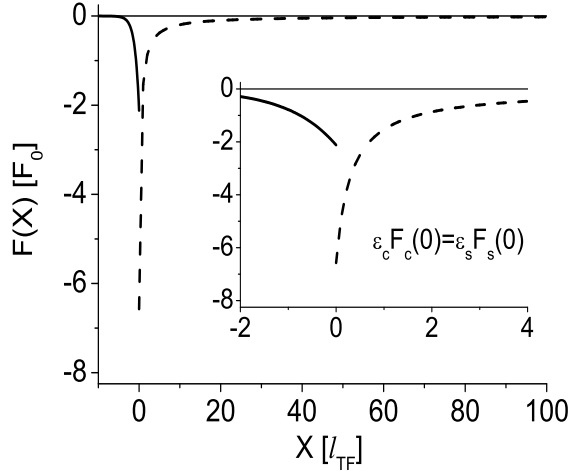}
\caption{{\itshape \small Distribution of the electric field $F$ in units of $F_0=\frac{kT}{el_{TF}}$ in equilibrium ($j=0$) and
barrier-free case ($\Delta=0$) as function of the coordinate $ X $ in units of $l_{TF}$. The inset shows the discontinuity of the
electric field at the interface.}} \label{infequid0}
\end{center}
\end{figure}

The solution for the dimensionless electric field is shown in Fig.\ref{infequid0}, where barrier-free injection ($\Delta=0$) from
an ITO electrode in an organic semiconductor is considered.
The solution can be interpreted as follows. Holes diffuse from the electrode into the organic semiconductor. This results in a
negative electric field causing a drift current opposite but equal in absolute value to the diffusion current, so that the net
current is zero. As a consequence, a space-charge zone is established in the electrode and in the organic semiconductor, which
in total is neutral. This space-charge region is very thin in the electrode
but extends far into the organic semiconductor. The weak decay of the electric field ($|F_s( X )|\sim 1/ X $) in the organic
semiconductor is due to a missing charge-carrier background in the insulator.
The distribution of the electric field changes when a non-vanishing injection barrier is introduced. Due to an impeded charge
carrier injection, the electric field is solely reduced close to the interface, leaving the field far in the organic material
invariant.

Depending on the material parameters chosen, the parameter $\alpha$, Eq.(\ref{alpha}), may be very large or very small
(here: $\alpha=5.6\cdot 10^{-3}$). In these limits the nonlinear equation (\ref{bcequi}) can be solved approximately.
If $\alpha<<1$, two characteristic regions arise depending on the relation between $\Delta$ and $\Delta_0=kT\ln(2/\alpha)$
(here: $\Delta_0\cong 0.15\mbox{ eV}$). If the barrier height is so small that $\Delta<<\Delta_0$, $F_s(0)$ is given by:
\begin{equation}
F_s(0)=-(A\,\epsilon_c/\epsilon_s) \left[ 1-(1+A/2)^{-1}\ln(A\,\epsilon_c/\epsilon_s) \right],
\label{Fnullappr1}
\end{equation}
with $A=\ln(2/\alpha)-\Delta/kT>>1$. If the barrier is so large that $\Delta>>\Delta_0$ then
\begin{equation}
F_s(0)=-\sqrt{\frac{2}{\alpha}} \exp \left ( -\frac{\Delta}{2kT} \right ).
\label{Fnullappr2}
\end{equation}
The latter relation is also valid for $\alpha>>1$ and arbitrary $\Delta$s. Eq.(\ref{Fnullappr2}) demonstrates the exponential
suppression of the electric field within the space-charge region by the injection barrier, since less charge carriers are transferred
across the contact.

The integration of the electric field leads to a voltage $V_{equi}$ to maintain equilibrium. The equilibrium voltage $V_{equi}$
is often referred to as the contact potential \cite{Sze}. Since the equilibrium field decreases with $1/X$,
the integration over the entire infinite semi-space diverges so that a cut-off length $L$ has to be introduced. This results from the
fact that a perfect insulator excluding any intrinsic charge carriers is considered. The analytical
expression for $V_{equi}$ reads,
\begin{equation}
V_{equi}=\frac{kT}{e} \left[ \frac{\epsilon_s}{\epsilon_c} F_s(0) - 2 \ln\left( 1-\frac{F_s(0)L}{2l_{TF}} \right)\right],
\label{Vequi}
\end{equation}
\noindent where $F_s(0)$ results from the boundary condition in equilibrium, Eq. (\ref{bcequi}). The existence of such an equilibrium
voltage has practical consequences for organic electronic devices. In organic photovoltaic cells, the contact potential
is known to reduce the open-circuit voltage considerably once the barrier height is small. In bulk-
heterojunction solar cells based on [6,6]-phenyl C$_{61}$-butyric acid methyl ester (PCBM) as electron acceptor and
poly[2-methoxy-5-(3',7'-dimethyloctyloxy)-p-phenylene vinylene] (OC$_1$C$_{10}$-PPV) as electron donator sandwiched between
ITO/PEDOT:PSS and LiF/Al electrodes a reduction of the open circuit voltage of $0.4\mbox{ eV}$ is observed
\cite{Mihailetchi2003}.

\subsection{Steady state}
Now, the steady state situation, where a constant current $j$ flows across the ITO/organic semiconductor system, will be
considered. The solution is given by Eq.(\ref{solspecial}) and the unknown constant $C_1$ has to be found numerically from
Eq.(\ref{bcssx0}).

\begin{figure}[h]
\begin{center}
\includegraphics[height=6.4cm,width=7.5cm]{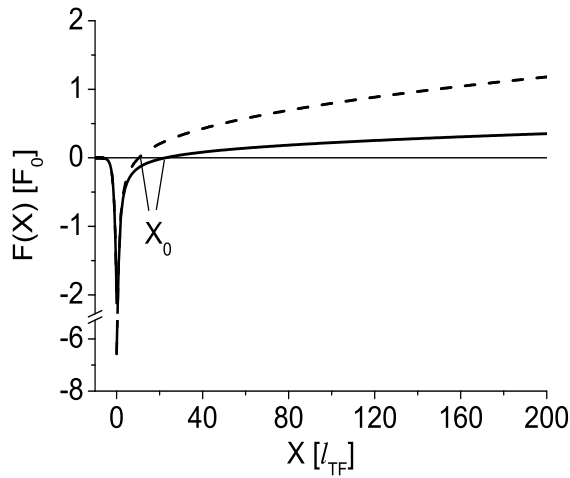}
\caption{{\itshape \small Distribution of the electric field $F$ in units of $F_0=\frac{kT}{el_{TF}}$ for barrier-free
charge-carrier injection and a constant current of $j=10\mbox{ mA}/\mbox{cm}^2$ (solid line) and $j=100\mbox{ mA}/\mbox{cm}^2$
(dashed line) as function of the coordinate $x$ in units of $l_{TF}$. }}
\label{infssd0}
\end{center}
\end{figure}

In Fig.\ref{infssd0}, the solution for the dimensionless electric field $F$ is depicted for current densities of
$j=10,100\mbox{ mA}/\mbox{cm}^2$ and barrier-free injection ($\Delta=0$).

Comparison with the equilibrium solution shows that the distribution of the electric field is, near the electrode, virtually
independent of the net current density, but is substantially effected by the current deep in the dielectric bulk.
For $\Delta = 0$ many holes diffuse into the organic semiconductor ($F_s'(0)$ large) and a strong negative field in the vicinity
of the interface emerges. This negative field compensates for the diffusion into the organic semiconductor to such an extend that
the net current reaches $j$. However, far away from the contact where diffusion is negligible, the electric field has to be
positive (together with the current). Thus, there is a position $X = X _0$ where the electric field changes sign, as can be seen
in Fig.\ref{infssd0}. Since a vanishing electric field strength is assigned to the ohmic contact itself, the position $X _0$ is
often referred to as the virtual electrode \cite{Rose1955}.

\begin{figure}[h]
\begin{center}
\includegraphics[width=9.5cm]{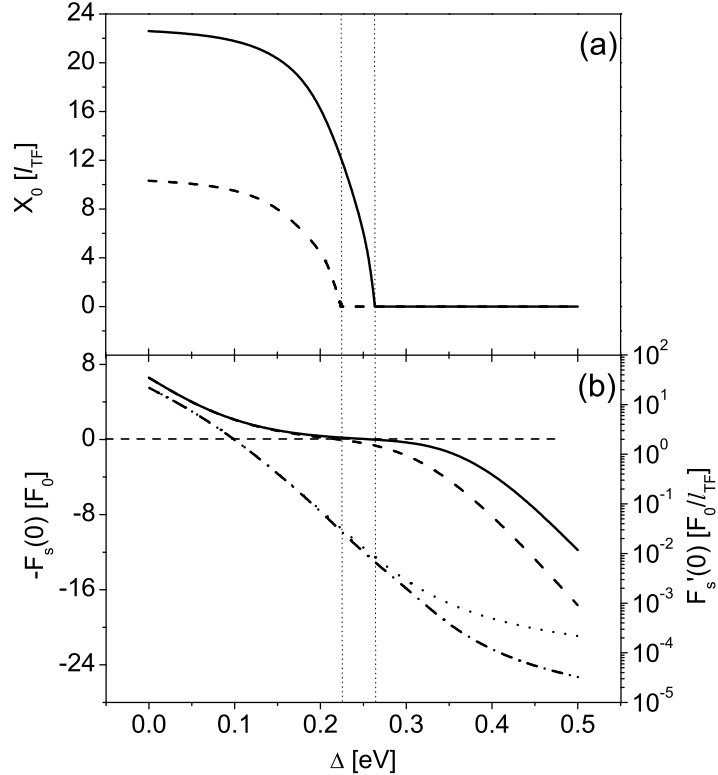}
\caption{{\itshape \small a) Position of the virtual electrode $ X_0$ in units of $l_{TF}$ for current densities of
$j=10\mbox{ mA}/\mbox{cm}^2$ (solid line) and $j=100\mbox{ mA}/\mbox{cm}^2$ (dashed line) as a function of barrier
height $\Delta$.
b) Electric field strength at $X=0$ inside the semiconductor for current densities of
$j=10\mbox{ mA}/\mbox{cm}^2$ (solid line) and $j=100\mbox{ mA}/\mbox{cm}^2$ (dashed line) as a function of barrier height
$\Delta$. The axis on the right hand side indicates the corresponding values for the derivative of the electric field for
current densities of $j=10\mbox{ mA}/\mbox{cm}^2$ (dashed-dotted line) and $j=100\mbox{ mA}/\mbox{cm}^2$ (dotted line). }}
\label{infE0}
\end{center}
\end{figure}

\begin{figure}[h]
\begin{center}
\includegraphics[height=6.4cm,width=7.5cm]{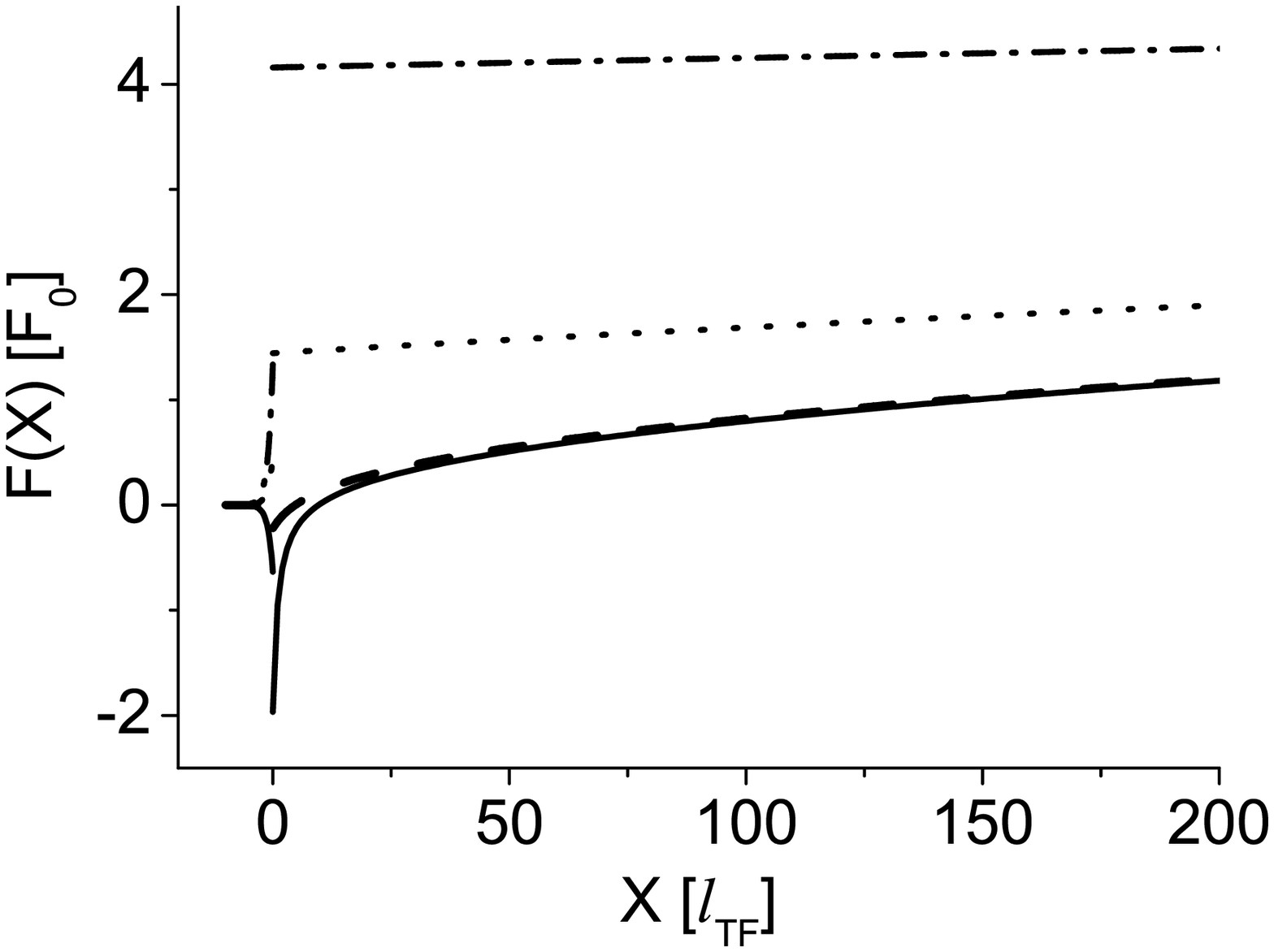}
\caption{{\itshape \small Distribution of the electric field $F$ in units of $F_0=\frac{kT}{el_{TF}}$ for a constant current
of $100\mbox{ mA}/\mbox{cm}^2$ and barrier height of $\Delta=0.1\mbox{ eV}$ (solid line), $\Delta=0.2\mbox{ eV}$ (dashed line),
$\Delta=0.3\mbox{ eV}$ (dotted line) and $\Delta=0.35\mbox{ eV}$ (dash-dotted line) as function of the coordinate $X$ in units
of $l_{TF}$. }}
\label{infssdmult}
\end{center}
\end{figure}

For a detailed discussion of the field distribution in the insulator, the dependence of the position of the virtual electrode,
the electric field at the contact $F_s(0)$, and the charge-carrier density (represented by $F'_s(0)$) are shown in Fig.\ref{infE0}
as functions of $\Delta$ and for two typical current densities. Additionally, the distribution of the electric field is depicted
in Fig.\ref{infssdmult} for different injection barriers and a current density of $j=100\mbox{ mA}/\mbox{cm}^2$.

For small $\Delta$s, the electric field at the interface is negative, the charge-carrier density is large, and the virtual
electrode is far inside the insulator. A charge-carrier reservoir is formed in the near contact region and the current across
the residual insulator $X>>X_0$ is supplied by this reservoir. From Fig.\ref{infssd0} and \ref{infssdmult} it becomes evident
that the electric field follows $F\sim \sqrt{ X - X _0}$ for $X>>X_0$ and hence, resembles the field distribution of SCLC
assuming ohmic boundary conditions \cite{Lambert}.

Increasing the injection barrier, the position of the virtual electrode is hardly effected until the injection barrier
exceeds  $ 0.1 \mbox{ eV}$. For a further increased barrier it approaches the physical electrode rapidly. Simultaneously,
$F_s(0)$ approaches zero and is pinned there for quite a wide $\Delta$-range. However, this does not mean that the ohmic boundary
conditions ($F_s(0)=0$ and $p_s(0)\rightarrow \infty$) assigned to an ideal contact \cite{Lambert} is a good approximation in this
case. In fact, the density of the charge-carriers at $X=0$ is relatively low, though $F_s(0)=0$ holds. $F_s( X )$ reflects neither
$\sqrt{ X - X_0}$ nor a constant form.

By further increasing $\Delta$, the amount of injected charge carriers remains small, $X_0$ is located at the physical electrode
and $F_s(0)$ becomes positive, while the field in the insulator is weakly dependent on $X$.
The charge-carrier reservoir in the insulator is depleted and due to the few charge-carriers at the contact, a strong positive
field is required to drive the current across the near contact region.

Note that for $F_s(0)<0$ (or $X_0>0$),
the position of the virtual electrode depends strongly on the induced current density while $F_s(0)$ and $F'_s(0)$
do not, reflecting the fact that close to the contact the equilibrium field distribution is hardly effected by the current but
further in the insulator it is. For barrier-free injection, the virtual electrode is shifted from approximately $X_0=25 \mbox{ nm}$
to $10 \mbox{ nm}$ once the current increases from $j=10\mbox{ mA}/\mbox{cm}^2$ to $j=100\mbox{ mA}/\mbox{cm}^2$. This is illustrated
in Fig. \ref{infssd0} and \ref{infE0}. Vice versa, the influence of the current density on the field at the contact becomes
recognizable for larger barriers, once the charge carrier reservoir is depleted and the virtual electrode coincides with the
physical one. This is due to the fact that a strongly increased positive field is required to support an additional current
density, since the charge-carrier density is small.

From Fig.\ref{infE0} it can be seen that the minimal injection barrier required to result in a match of virtual and physical
electrode ($F_s(0)=0$) is shifted to smaller barriers for higher current densities. By means of the solution~(\ref{solspecial}),
this may be formulated as an exact relation between the quantities involved. The requirement $F_s(0)=0$ results in the particular
value of the constant $C_1=z_0 (2/|\iota|)^{1/3}$ where $z_0\simeq -1.02$ is the first zero of the Airy function
$\mathrm{Ai}^{\prime}(z)$. Using Eq.(\ref{bcssx0}), the current magnitude $\iota_0$, which suppresses the electric field at the
interface, is determined to
\begin{equation}
\iota_0=\frac{\exp(-3\Delta/2kT)}{2^{1/2}(\alpha |z_0|)^{3/2}}.
\label{fieldnode}
\end{equation}
From the above equation it becomes evident that for increased injection barriers, the minimal current density required to obtain
a match of virtual and physical electrode is exponentially reduced.

Knowledge about the distribution of the electric field gives access to the voltage drop $V$ across the system for a given current
density $j$ and hence, to its IV-characteristics. As the model consists of infinite semi-spaces, integration over the electric field
diverges when being carried out over the entire space, i.e. from $X=-\infty$ to $X=+\infty$. This holds in equilibrium as well as
in steady state. However, since the focus of this work is to analyze the contact phenomena arising from the conductor/insulator
junction, two simplifications are introduced. Firstly, according to the definition of the contact potential the voltage drop is
calculated by introducing a cut-off length $L$ corresponding to the typical thickness of the organic layer. Hence, the bulk
conduction in an organic layer with finite thickness is taken into account. Secondly, since the voltage drop across a conductor
bulk of finite thickness is small due to its high conductivity, the constant part equal to the asymptotic constant field value
times the macroscopic conductor thickness is subtracted from the integral over the conductor. Then, the voltage drop over the
entire system reads:
\begin{equation}
V=\frac{kT}{e} \left\{
\frac{\epsilon_s}{\epsilon_c} F_s(0) - \iota \alpha \frac{\mu_s}{\mu_c} \frac{{\cal N}}{p_\infty}
- 2\ln{
\left|
\frac{\mathrm{Ai}\left[(\iota/2)^{1/3}(L/l_{TF}+C_1)\right]}
{\mathrm{Ai}\left[(\iota/2)^{1/3}C_1\right]}
\right|}
\right\}
- V_{equi}.
\label{voltinf}
\end{equation}
\noindent Here, the first two terms in braces represent the voltage drop in the electrode. The third term in braces results from
the voltage drop within a distance $L$ from the contact inside the organic semi-space and the last term corrects the voltage
by its equilibrium value, Eq.(\ref{Vequi}). The two (not independent) constants $F_s(0)$ and $C_1$ result from the steady-state
boundary condition, Eq.~(\ref{bcssx0}).

\begin{figure}[h]
\begin{center}
\includegraphics[height=6.4cm,width=7.5cm]{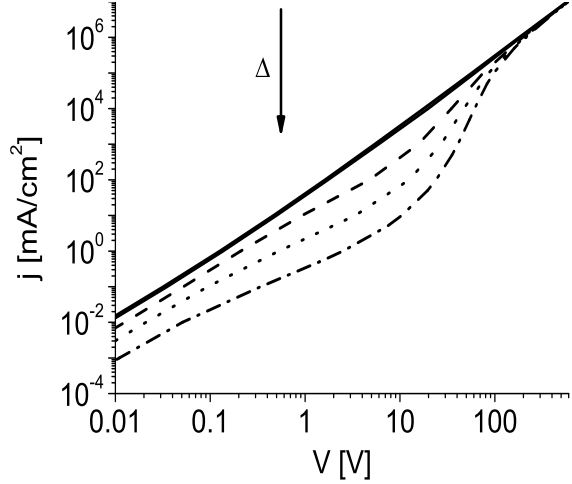}
\caption{{\itshape \small IV-characteristics for barrier height of $\Delta=0;0.2;0.3;0.35;0.4\mbox{ eV}$.
The curves for $\Delta=0$ and $\Delta=0.2 \mbox{ eV}$ nearly merge in the above representation.}}
\label{infivdmult}
\end{center}
\end{figure}

In Fig.\ref{infivdmult}, the resulting IV-characteristics for different barrier heights are presented assuming $L=100\,\mbox{nm}$.
In the displayed voltage region, the barrier-free contact ($\Delta=0$) is able to supply more charge-carriers than the bulk of the
organic semiconductor can transport. Hence, the entire system appears to be space-charge limited with a current approximately
$j\sim V^2$ for all voltages. Deviations from the Mott-Gurney law are due to the formation of the space-charge region emerging by charge-carrier diffusion
at $X<X_0$. As the current increases, the position of the virtual electrode approaches the real electrode and the width of this space charge region
decreases. Thus, the calculated curve reproduces the Mott-Gurney law \cite{Lambert} more exactly for higher voltages.

In the presence of non-vanishing barriers, the current is substantially reduced. In the low voltage regime this reduction is about
one order of magnitude once the height of the barrier increases from $\Delta=0$ to $\Delta=0.4\mbox{ eV}$. For high $\Delta$s ,
one observes a transition from a $j\sim V^2$ to a $j\sim V$ dependence, followed by an exponential current increase with the
applied voltage. Since the SCLC represents an upper limit for the current flow through the system, all IV-curves approach
the SCLC regime for even higher voltages.

\begin{figure}[h]
\begin{center}
\includegraphics[height=6.4cm,width=7.5cm]{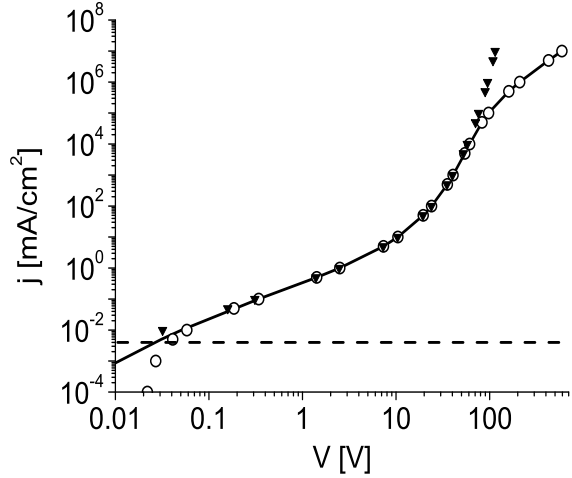}
\caption{{\itshape \small IV-characteristic for barrier height of $\Delta=0.4\mbox{ eV}$. The solid line has been obtained
using the exact solution (\ref{solspecial}) of the complete drift diffusion equation (\ref{dglorgdimless}). The circles show
the IV-characteristic calculated with the help of the
semi-analytical approach derived for high injection barriers, Eqs.(\ref{driftfield})-(\ref{approxcurrent}), and the triangles
display the injection current $j_{inj}$ calculated with the help of Eq.(\ref{injeccurrent}). The dashed line shows a characteristic
current of the virtual electrode appearance.}}
\label{infivapprox}
\end{center}
\end{figure}

The calculated IV-characteristics for high injection barriers can be understood by considering some simple approximations.
Once the electric field at the interface becomes positive, the charge-carrier reservoir established in the organic semiconductor
is exhausted and no virtual electrode is present. Hence, diffusion is negligible all over the space and an electrical current is
virtually due to charge carrier drift only,
\begin{equation}
F_s'(X)F_s(X)=\iota.
\label{drift}
\end{equation}
The solution depends on $F_s(0)$ and reads \cite{Emtage1966},
\begin{equation}
F_s(X)=\sqrt{F_s^2(0)+2\iota X }.
\label{driftfield}
\end{equation}
Analyzing the drift equation (\ref{drift}) at $X=0$ by using the boundary condition, Eq.(\ref{bcssx0}) leads to an exponential
dependence of the dimensionless current $\iota$ on the local value of the electric field
\begin{equation}
\iota=\frac{F_s(0)}{\alpha} \exp \left ( \frac{\epsilon_s}{\epsilon_c} F_s(0)-\frac{\Delta}{kT}\right ) ,
\label{approxcurrent}
\end{equation}
where the right hand side of the boundary condition, Eq.(\ref{bcssx0}), has been neglected.
Solving Eq.(\ref{approxcurrent}) numerically with respect to $F_s(0)$ for a given current density $\iota$  determines the electric
field distribution, Eq.(\ref{driftfield}), and integrating the field over the entire system leads to the IV-characteristic.
This approximation offers a simple way to calculate IV-characteristics \cite{Blom2000} displayed exemplary for a
barrier height of $\Delta=0.4\mbox{ eV}$ in Fig.\ref{infivapprox}.

The agreement with the IV-characteristic calculated using the exact solution, Eq.(\ref{solspecial}), is perfect for all voltages
where diffusive transport of charge carriers is negligible. This includes the SCLC regime at high bias. Only at low voltages, where
the charge carrier reservoir in the organic semiconductor is not exhausted, deviations are recognizable. According to the exact
relation, Eq.~(\ref{fieldnode}), one can see that below a characteristic current $\iota\simeq 4\cdot 10^{-3}$ the virtual
electrode at $X_0>0$ appears and the diffusion becomes important.

For a purely injection-limited regime space charge effects are of no importance, resulting in a constant electric field
all over the semiconductor semi-space. This field
coincides with the electric field strength at the interface. In such a case $F_s(0) \cong V e l_{TF}/ kT L$ and
Eq.(\ref{approxcurrent}) governs the injection limited IV-characteristics alone,
\begin{equation}
j_{inj}=e\mu_s \frac{V}{L} {\cal N} \exp \left [ -\frac{\Delta}{kT} + \frac{e \epsilon_s l_{TF} V}{\epsilon_c kTL}  \right ].
\label{injeccurrent}
\end{equation}
Eq.(\ref{injeccurrent}) resembles
the result of the drift-diffusion equation \cite{Sze, Blom2000, Crowell1966}. Yet, the self-consistent treatment of the injection problem yields directly
a barrier lowering arising from the potential energy a charge carrier gains
on the conductor side of the interface. In contrast to the $\ln(j)\propto \sqrt V$ dependence of the Schottky-lowering predicted from
the image-charge potential in the single-electron picture, the current density depends exponentially on
the external voltage $V$. The injection-limited current $j_{inj}(V)$ is depicted in Fig.\ref{infivapprox}.

Once space charge effects become predominant, $F_s(0)$ can be neglected in Eq.(\ref{driftfield})
and the temperature-independent Mott-Gurney law is reproduced. This occurs as soon as the field-induced barrier lowering has
proceeded to such extent that the contact can again establish space charge in the organic semiconductor. Hence, a transition
from injection-limited to space-charge-limited current is observed. As may be seen from Fig.\ref{infivapprox}, the crossover
from injection-limited to SCLC occurs at rather high voltages in presence of a medium injection barrier of $0.4 \mbox{ eV}$.
This has its origin
in the weak barrier lowering for increasing external voltage. Due to the low screening length in the conductor, a high voltage
drop across the entire system is required to result in an energy gain in the conductor being comparable to the injection-barrier
energy. Note that in contrast to a space-charge formation resulting from diffusion, the electric field at the contact is positive
once the space-charge formation is solely due to a drift-controlled injection. Such a SCLC regime occurs also for barrier-free
injection as soon as the diffusive filling of the space-charge region is weaker than its depletion due to the external applied
field. This, however, requires a very high bias. As a result, the SCLC is directly supplied by an efficient charge-carrier drift
out of the conductor.

From Eq.(\ref{injeccurrent}) and Mott-Gurney law one can estimate the upper limit of the barrier required to retain SCLC over
the whole voltage range:
\begin{equation}
\Delta_{crit}=kT + kT \ln \left ( \frac{16}{27}\frac{L}{l_{TF}}\frac{{\cal N}}{p_\infty}\frac{\kappa_\infty}{kT}\right ).
\label{deltacrit}
\end{equation}
Assuming the parameters introduced before, we can predict a critical barrier height of $0.25 \mbox{ eV}$. This is in slight
contradiction to device models in which a Schottky-type barrier lowering is assumed \cite{Scott1999,Malliaras1999,Campbell1998}.
Here, one would expect the crossover between SCLC and injection-limited current to occur at an injection barrier of
$\Delta_{crit}\simeq0.35\mbox{ eV}$. However, up to now the experimental data available predict the crossover to be at a
barrier height between $0.2\mbox{ eV}$ and $0.35\mbox{ eV}$, being consistent with both approaches.

\section{Conclusions}

In the calculation of the charge-carrier transport through
insulators, a fundamental question about the boundary conditions
generally arises when a charge-carrier injecting interface has to be
involved. Typically, boundary conditions at the interface are
chosen, fixing there the charge carrier density and/or the electric
field. However, at the conductor/insulator contact the system is
ill-defined, meaning that especially at the interface the charge
carrier density and the electric field strongly depend on the
condition of the system.

In this paper, a one-dimensional analytical
model describing the charge-carrier transport across a
conductor/insulator junction was presented, where boundary
conditions are defined far into the conductor and the insulator,
respectively. Here, the influence of the two materials on each other
is negligible so that they can be regarded as independent.
Considering the Poisson equation and assuming the electric
displacement as well as the electrochemical potential being
continuous over the entire system, the electric field distribution
and the current-voltage characteristic were derived. The model
predicts SCLC, injection-limited conduction as well as the crossover
between the two limits. In most current regimes, the influence of
the self-consistent treatment is noticeable. For pure
injection-limited conduction, an injection current similar to the
prediction of the drift-diffusion theory was derived. However, due
to the consistent treatment of the injection problem in one
dimension, a barrier lowering different to the one predicted from
the three-dimensional image-charge potential in the single-electron
picture emerges. This injection-barrier lowering results from the
potential energy gain (or loss) of the charge-carriers in the
electrode. Also for SCLC, a deviation of the Mott-Gurney law is
observable as long as the virtual electrode does not match the
physical one. Only at high bias the self-consistent treatment has no
influence on the IV-characteristic and the well known Mott-Gurney
law is fully reproduced.

\section{Acknowledgements}
Useful discussion with R. Schmechel is gratefully acknowledged. The authors would express their thanks to the Deutsche
Forschungsgemeinschaft for financial support of the Sonderforschungsbereich 595. The coauthors dedicate this article 
to the memory of Frederick Neumann whose life was taken by a tragic accident.

\end{document}